\documentclass[11pt,a4paper]{article}
\usepackage[makeroom]{cancel}
\usepackage{bm}
\usepackage{amsmath}
\usepackage{amsfonts}
\usepackage{graphicx}
\usepackage{nicefrac}
\usepackage{authblk}
\usepackage{cite}
\usepackage[left=2.54cm,top=2.54cm,right=2.54cm,bottom=2.54cm,nohead]{geometry}
\DeclareGraphicsExtensions{.png,.jpg,.pdf}
\usepackage{epstopdf}
\usepackage{float}
\usepackage{subfigure}
\usepackage{fouriernc} 
\usepackage{enumitem}

\title{Theory of water desalination with intercalation materials}
\usepackage{authblk}
\makeatletter
\renewcommand\AB@authnote[1]{}

\makeatother

\author[1,2]{K. Singh,\textsuperscript{1,2}}
\author[3]{H.J.M. Bouwmeester,\textsuperscript{3}}
\author[1,2]{L.C.P.M de Smet,\textsuperscript{1,2}\\}
\author[4,5]{M.Z. Bazant,\textsuperscript{4,5}} 
\author[2]{P.M. Biesheuvel\textsuperscript{2}}
 
\affil[1]{Laboratory of Organic Chemistry, Wageningen University, The Netherlands.}
\affil[2]{Wetsus, European Centre of Excellence for Sustainable Water Technology, %Oostergoweg 9, 8911 MA 
Leeuwarden, The Netherlands.}
\affil[3]{Electrochemistry Research Group, Membrane Science and Technology, Department of Science and Technology, MESA+ Institute for Nanotechnology, University of Twente, % Drienerlolaan 5, 7522 NB Enschede, 
The Netherlands.}
\affil[4]{Department of Chemical Engineering,  Massachusetts Institute of Technology, USA.}
\affil[5]{Department of Mathematics,  Massachusetts Institute of Technology, USA.}

\date{} %remove date

\begin{document}

%\pagenumbering{roman}

\renewcommand{\t}{\widetilde}
\renewcommand{\t}{}
\newcommand{\s}[1]{\mathrm{_{#1}}}

\maketitle

\begin{abstract}
We present porous electrode theory for capacitive deionization (CDI) with electrodes containing nanoparticles that consist of a redox-active intercalation material. %Ionic transport in the aqueous phase is assumed to be rate-limiting. 
A geometry of a desalination cell is considered which consists of two porous electrodes, two flow channels and an anion-exchange membrane, and we use Nernst-Planck theory to describe ion transport in the aqueous phase in all these layers. A single-salt solution is considered, with unequal diffusion coefficients for anions and cations. Similar to previous models for CDI and electrodialysis, we solve the dynamic two-dimensional equations by assuming that flow of water, and thus the advection of ions, is zero in the electrode, and in the flow channel only occurs in the direction along the electrode and membrane. In all layers, diffusion and migration are only considered in the direction perpendicular to the flow of water. Electronic as well as ionic transport limitations within the nanoparticles are neglected, and instead the Frumkin isotherm (or regular solution model) %Langmuir-Temkin equation 
is used to describe local chemical equilibrium of cations between the nanoparticles and the adjacent electrolyte, as a function of the electrode potential. Our model describes the dynamics of key parameters of the CDI process with intercalation electrodes, such as effluent salt concentration, the distribution of intercalated ions, cell voltage, and energy consumption. 
 
\end{abstract}

\section{Introduction} \label{sec:Intro}

Capacitive Deionization (CDI) is a method of water desalination where two porous electrodes adsorb ions from the aqueous phase, and later release the ions again~\cite{Suss_2015,Biesheuvel_2017,He_2018}. Salt removal is achieved by alternatingly charging and discharging the electrodes which are connected through an electrical circuit. The porous electrodes typically contain three phases: water-filled pores for ion transport, a conducting material for electronic current, and a phase where ions are temporarily stored, see Fig.~1. Most work on CDI uses electrodes with ion storage in materials based on carbon (activated carbon, carbon nanotubes, graphene, etc.) where ions are stored in the electrical double layer (EDLs) along the carbon surface~\cite{Huang_2017}. 

Another class of materials for ion storage is based on intercalation host compounds, which are examples of redox-active materials that are receiving increasing attention because of their potential for higher salt adsorption, lower energy consumption and tunable ion selectivity.  Although our general theory below could be applied to arbitrary intercalation compounds, adsorbing cations or anions, we will focus on the more common case of cation intercalation.  Examples of such materials are nickel hexacyanoferrate (abbreviated as NiHCF, with chemical structure NiFe(CN)$_6$), which is a Prussian Blue analogue (PBA)~\cite{Rassat_1999,Itaya_1986,McCargar_Neff,Wessells_2011,
Porada_Smith,Kim_2017,Ma_2017}, sodium manganese oxide (NMO)~\cite{Pasta_2012,Lee_2014,Smith_Dmello,Shanbhag_2017,Wu_2017}, and iron or titanium phosphates~\cite{Kim_2016,Shanbhag_2017} (which are also used in Li-ion batteries). In response to an applied voltage, cations intercalate, or are reversibly inserted, into the %nanoscale ``pores'' (molecular channels) within the crystal of 
these redox-active host materials.  
The host crystal structure contains transition metal atoms, such as Fe or Mn which change redox-state upon injection of electronic charge. When this material is charged more negatively (redox atoms in the crystal are reduced), to maintain electroneutrality extra cations are incorporated in the pores of the crystal, thus desalinating the water outside the particles. The intercalation materials mentioned above are highly cation selective (not incorporating anions), since %they maintain strong negative charge in 
the crystal is negatively charged, which provides a background counter-charge for the intercalated cations.%, which do not significantly change their redox state as they move through the crystal. 

\begin{figure}[ht]
\centering
\includegraphics[width=0.8\textwidth]{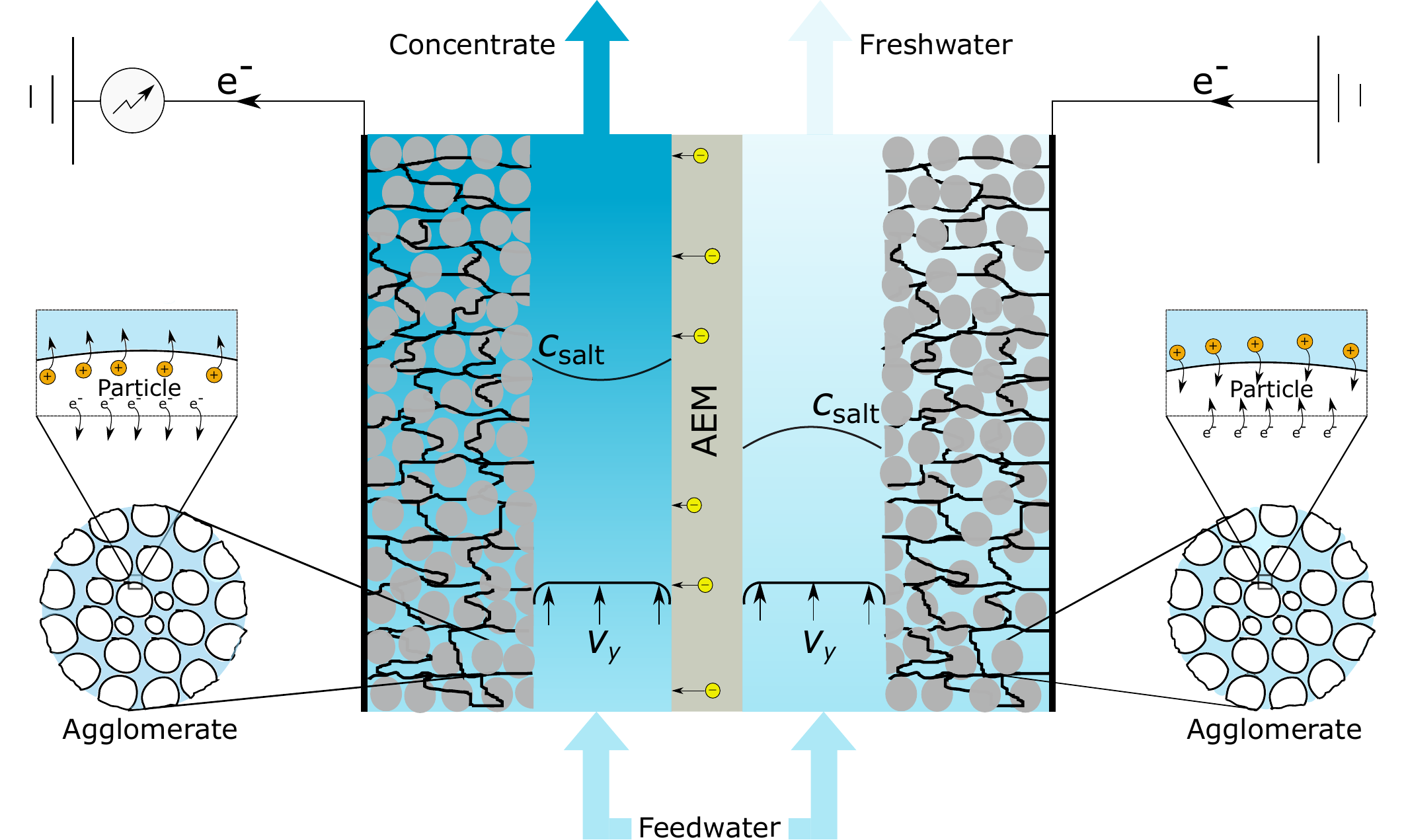}
\caption{Schematic of the CDI cell with porous electrodes containing nanoparticles of redox-active intercalation material. Electronic connections through the electrode (drawn as black lines) allow for injection of charge from an external circuit. Nanoparticles are aggregated into water-filled agglomerates. The electrodes are separated by a pair of spacer channels for water flow, and an anion-exchange membrane which preferentially allows transport of anions. The schematic captures the moment when ionic current flows left to right leading to de-intercalation of Na$^+$ on the left, and intercalation on the right. Depicted in the spacer channels are typical profiles in salt concentration and velocity of the water.}
\label{fig:schematic_cell}
\end{figure}

To describe desalination by traditional CDI based on double-layer charging, porous electrode theory has been used since the 1970s \cite{Johnson_Newman,Ferguson_2012}. Johnson and Newman developed the first such theory for CDI and made the assumption that for each electron injected in the EDL of a carbon matrix, one counterion adsorbs~\cite{Johnson_Newman}. Going beyond this approximation, Bazant {et al.} initiated the microscopic theory of double-layer charging dynamics and salt removal for the model problem of parallel-plate blocking electrodes~\cite{Bazant_2004}. Since 2010, Biesheuvel and Bazant incorporated this physics into a porous electrode theory for double-layer CDI, applied to carbon-based materials~\cite{Biesheuvel_Bazant_2010}, which allows for incorporation of any quasi-equilibrium, thin or thick, EDL structure. This generalized theory can be used for EDL models that consider both counterions and co-ions, ion mixtures of arbitrary valency, presence of chemical surface charge, and Faradaic charge transfer~\cite{Biesheuvel_2012}. 

Porous electrode theories have also been developed for intercalation materials, albeit motivated by applications to Li-ion batteries rather than CDI.  In 1982, West {et al}.~\cite{West_1982} presented %an elegant 
porous electrode theory for energy storage in intercalation, or insertion, materials, where Nernst-Planck theory describes both ion transport across the porous electrode and inside the intercalation material. At the interface of electrolyte and  intercalation material, the Frumkin intercalation isotherm %Langmuir-Temkin equation 
is used~\cite{Conway_Gileadi,Levi_Aurbach}, which provides a relation between electrode potential (relative to the electrolyte phase), ion concentration in solution, and intercalation degree (the degree to which the intercalation material is filled with cations). This isotherm is an example of more general regular solution theory describing the homogeneous Gibbs free energy of mixing of two or more constituents with finite size (e.g., mean field approximation on a crystal lattice), as used in phase-field models of intercalation materials~\cite{Nauman_He,Bazant_2013,Lai_2010}. Regular solution models can be used to describe the thermodynamics of host materials containing multiple mobile ionic species and point defects, such as vacancies and interstitials, and form the basis for simulations of nonlinear diffusion, phase separation, and electrochemical reactions~\cite{Bazant_2013}, in both single-crystal nanoparticles and multiphase porous electrodes~\cite{Ferguson_2012,Smith_Bazant_2017}.

In the present work we set up and solve a model for CDI with intercalation materials along the lines of Johnson and Newman~\cite{Johnson_Newman} and West {et al.}~\cite{West_1982}. We simplify the model of West {et al.} by neglecting  transport limitation within the intercalation material, a common approximation in the modeling of  battery electrodes consisting of intercalation nanoparticles~\cite{Ferguson_2014,Smith_Bazant_2017}. This is a valid assumption when the nanoparticles are very small and accessible on all sides by the ions in the electrolyte (water) with fast insertion reactions~\cite{Ferguson_2012}. A rigorous criterion is not available for validity of this assumption in all situations, because ion transport in the electrode and adsorption in the particles is not a simple steady-state resistances-in-series process. But a first comparison can be made on the basis of a ``critical flux'' of Na$^+$-ions in the two phases, $J_\text{crit}=cD/L$, with $c$ a typical concentration, $D$ a typical diffusion coefficient and $L$ a typical distance. For the electrolyte we have $c=10$ mM, $D=2.2\times 10^{-10}$ m$^2$/s (based on a porosity of 30$\%$ and tortuosity based on the Bruggeman equation), and $L=250$ $\mu$m, while for the nanoparticles, $c=2.2$ M (at $\vartheta=0.5$ for NiHCF, see ref.~\cite{Porada_Smith}), $D=6.9\times 10^{-16}$ m$^2$/s~\cite{You_2013}, and $L=5$~nm (based on a particle with a size of 30 nm; note that for a sphere, a typical diffusion depth is equal to the volume/area-ratio, which is on sixth of the size of the sphere). These numbers give a critical flux for Na$^+$ in the nanoparticles that is about 35$\times$ larger than in the electrolyte phase in the electrode, thus underpinning the choice to focus on transport in the electrolyte.

\begin{figure}[ht]
\centering
\includegraphics[width=0.9\textwidth]{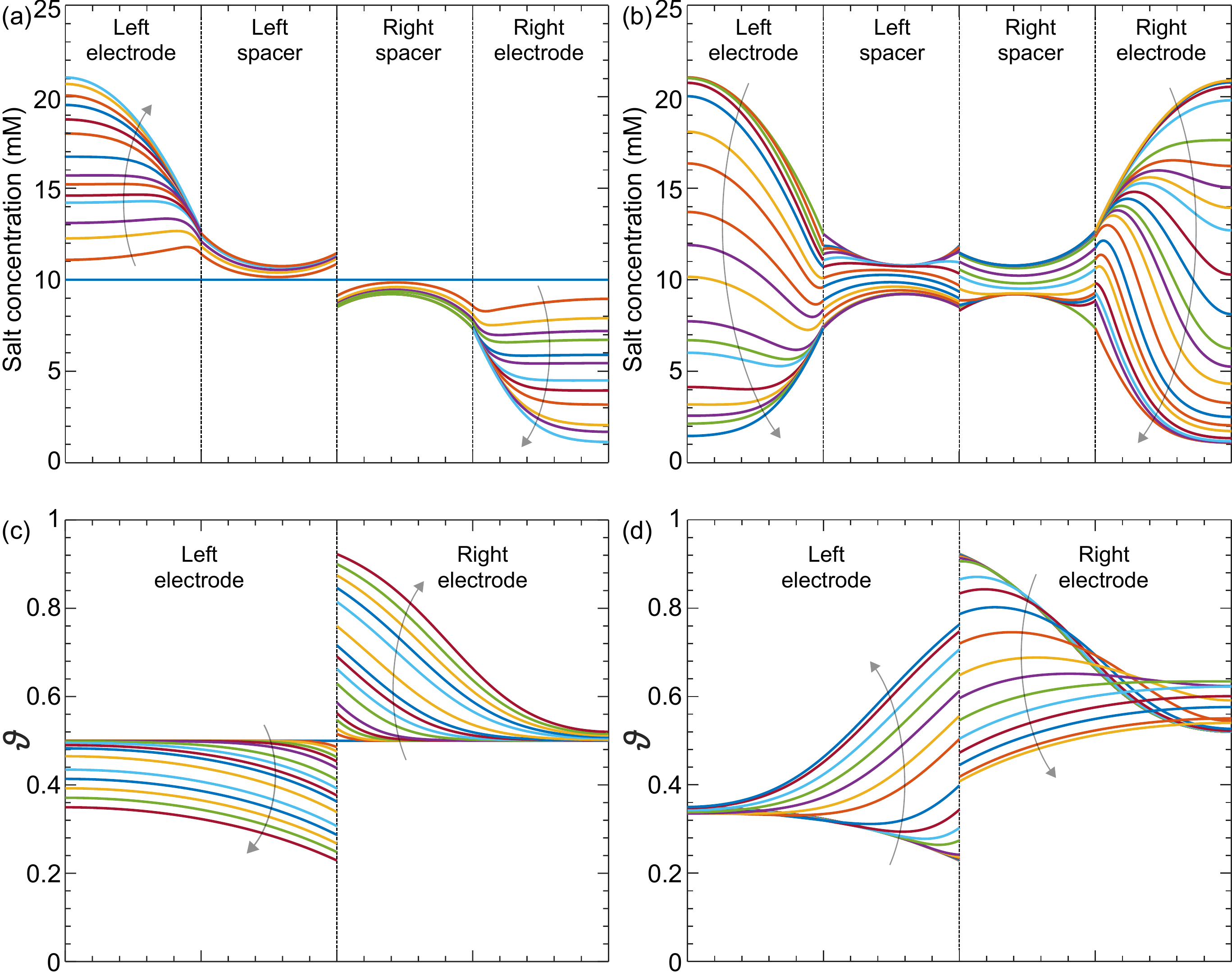}
\caption{Time-dependent profiles of salt concentration and intercalation degree during water desalination with intercalation electrodes. A,B) Salt concentration in electrodes and spacers; C,D) Intercalation degree in the two electrodes, $\vartheta$. These profiles are given as a function of position in the cell during A,C) first charging step, and B,D) first discharge step. Arrows depict the progression of time.}
\label{fig:conc_profile_charge_discharge}
\end{figure}

Though we simplify the model by West {et al.}~\cite{West_1982}, we also extend their approach (which is for a single electrode) by considering a full electrochemical cell consisting of two electrodes, two flow channels (also called spacers or spacer channels), and an anion-exchange membrane (AEM). The geometry is sketched in Fig. 1. This cell design was invented by Smith and Dmello~\cite{Smith_Dmello} to solve the problem of how to desalinate water using two electrodes made of intercalation materials that both are only capable of adsorbing cations. 

Not only does our porous electrode theory relate to that of West {et al.}~\cite{West_1982} and Ferguson and Bazant~\cite{Ferguson_2012,Ferguson_2014}, it is also similar in structure to that of Johnson and Newman~\cite{Johnson_Newman}, except that we do not assume a constant capacitance, but instead implement the Frumkin intercalation isotherm, which for several intercalation materials has been shown to accurately describe chemical equilibrium between electron-conducting phase, intercalation material, and electrolyte~\cite{Conway_Gileadi,Levi_Aurbach,Bazant_2013,Smith_Bazant_2017}. According to this equation, capacitance is high when the intercalation degree in the material is at an intermediate value ($\vartheta=0.5$), while it drops dramatically when the material either is almost saturated with ions ($\vartheta \rightarrow 1$) or almost empty ($\vartheta \rightarrow 0$). %In ref.~\cite{Levi_Aurbach}, this equation is referred to as the Frumkin intercalation isotherm.

The electrode model is combined with electro-diffusion of ions in the two flow channels and in the AEM (see geometry depicted in Fig. 1). For ion transport in the pores of the electrode, and in the flow channel, we implement that the ions have a different diffusion coefficient. Besides, we include a full description of the AEM, allowing both passage of cations and anions. Because the membrane has a large concentration of positive fixed charge, the flux of anions is much larger than of cations. We make calculations of complete charge-discharge cycles and show the development over time of key parameters such as cell voltage, effluent salinity, and the intercalation degree of the electrodes. 

Note that our theory considers the intercalation particles to fill up homogeneously, without phase separation inside the solid phase. This phase separation is not occurring for NiHCF in aqueous electrolyte, but does happen for other systems, such as Li-insertion in FePO$_4$--based electrodes in organic solvent, where Li-dense and Li-sparse regions are both found within the LiFePO$_4$-particles. Extensions to porous electrode theory that are required to include phase separation in the solid phase are introduced and reviewed in Refs.~\cite{Bazant_2013,Ferguson_2012,Smith_Bazant_2017}.

\begin{figure}[ht]
\centering
\includegraphics[width=0.9\textwidth]{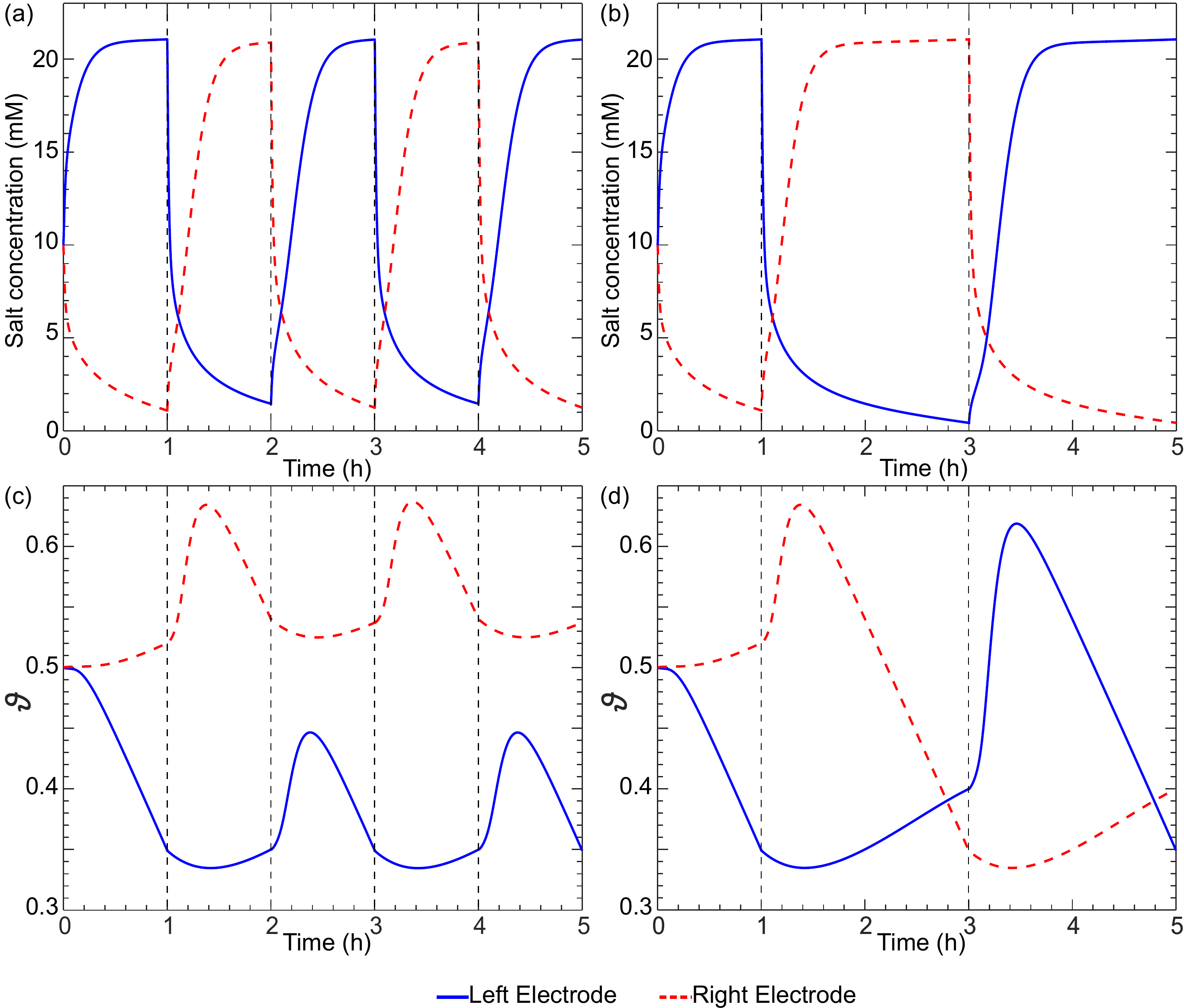}
\caption{Salt concentration and intercalation degree $\vartheta$ versus time for A,C) operation in Mode 1, and B,D) operation in Mode 2. In all cases calculation results are presented for the positions deepest in the electrode, thus furthest from the flow channel, for A,B) salt concentration, and C,D) intercalation degree.}
\label{fig:conc_theta_time_two_config}
\end{figure}

For an electrochemical cell based on porous electrodes containing nanoparticles of redox-active intercalation material, the kinetics of ion adsorption depends on various transport processes, and each must be included in a complete theory for the entire device. These processes include:
\begin{enumerate}%[leftmargin=*]
\vspace{-2.5mm}
\setlength\itemsep{0\baselineskip}
\item{Diffusion and migration of ions through the electrolyte-filled pores of the electrode;}
\item{Ion insertion (intercalation) into the redox-active material, and transport across these particles;}
\item{Electronic charge transport across the electrode, from an external current source all the way down into each nanoparticle;}  
\item{Ion transport (advection, diffusion and migration) in a transport channel located outside the electrode; and}
\item{Transfer of ions across the ion-exchange membrane.}
\end{enumerate}
\vspace{-2.5mm}
In the present work, transport by processes 1, 4 and 5 is addressed, while instead of addressing items 2 and 3 in detail, the transport resistance related to these processes is assumed to be infinitely low. Note that we do not consider a rate limitation in the intercalation step of ion transfer between the electrolyte and the intercalation materials. In many situations, this ion-transfer reaction may not be rate-limiting, %since the ion does not change solvation degree upon entry into the intercalation material, although there is an activation barrier for de-solvation and electron transfer to the compensating redox-active site in the crystal~\cite{Bai_2014}. Although we assume fast ion transfer here, the intercalation step could 
though it can be included in future extensions of our model, using various models of electrochemical reaction kinetics~\cite{Smith_Dmello,Bazant_2013,Smith_Bazant_2017,Bai_2014}.

Similar to Johnson and Newman~\cite{Johnson_Newman} and West {et al.}~\cite{West_1982}, we make use of the Nernst-Planck (NP) equation for ion transport, an equation which is valid for a sufficiently dilute electrolyte, and which describes each ion in the same way. This approach is mathematically different from models developed for (Li-ion) batteries, which incorporate a solution-phase potential as the driving force for transport of cations~\cite{Doyle_1993}.  This potential combines the electrical potential in solution with a term dependent on Li$^+$ ion concentration. Also thermodynamic non-idealities, beyond the NP-equation, are incorporated in such state-of-the-art battery models. Though these battery models were also successfully applied for CDI with intercalation materials~\cite{Smith_Dmello,Smith_2017}, nevertheless in the present work we follow refs.~\cite{Johnson_Newman,West_1982} and make use of porous electrode theory based on the NP equation. Advantages of the NP-approach are that it is mathematically easier to understand and implement, and in a later stage can be readily extended to ionic mixtures. Such modifications are not so straightforward to implement in the available codes for battery models~\cite{Smith_Bazant_2017}. For a dilute electrolyte, the NP-equation is a valid approximation, because thermodynamic non-idealities of the aqueous phase are not very pronounced. Thus we use the ideal NP equation for migration and diffusion of each ionic species in the different elements in the cell, extended with an advection-term for the flow channel. 

In summary, in the present work we aim to develop and present a mathematical model for water desalination by CDI using porous electrodes consisting of nanoparticles of intercalation material. Our approach is to use a theoretical framework which is not excessively complicated while it describes a large number of the physical and chemical processes that take place in CDI with intercalation materials. At a later stage, the present model can be extended in a straightforward manner to describe mixtures of salts, by using a generalized Frumkin intercalation isotherm for ionic mixtures~\cite{Porada_Smith,Erinmwingbovo_2017}. Such an extension would be very difficult to make in the classical theoretical framework of modeling Li-ion batteries that use quasi-electrostatic potentials and concentrated solution theory. In our present work we discuss two different modes of operation of the electrochemical cell and the influence thereof on key parameters relevant for understanding cell performance (e.g., salt concentration, intercalation degree, and cell voltage). 

\section{Theory} \label{s:theory}

We make a calculation for a complete CDI desalination cell which consists of two electrodes, two spacer channels, and an anion-exchange membrane (AEM) which is placed in the middle of the cell, see Fig.~\ref{fig:schematic_cell}. This arrangement, with the electrodes connected to an external electronic circuit, forms the desalination system that we will describe. The feed streams enter the spacer channels on one side and leave on the other, in a direction longitudinal to membrane and electrode. The spacer channels are in contact with an electrode on one side and the AEM on the other. The channels are not completely open, but contain a porous layer with a certain porosity, $\varepsilon_\text{s}$, and tortuosity, $\tau_\text{s}$. Feedwater is used with a sufficiently low  salt concentration for the assumptions of dilute solution theory to hold (activity coefficient, $\gamma\sim 1$)~\cite{Ferguson_2012}. Ion transport in these channels is described by the Nernst-Planck (NP) equation, extended to include advective flow, 
\begin{equation} \label{eq:nernst_planck_vector}
\bm{J}_{i}=c_{i}{{v}} - \frac{\varepsilon_\text{s}}{\tau_\text{s}}D_{i}\left(\nabla c_{i}- z_{i}c_{i}\nabla{\phi}\right)
\end{equation}
where $\bm{J}_{i}$ is the flux of ion $i$, ${{v}}$ is fluid velocity, and $c_{i}$ is ion concentration. We relate the tortuosity of the porous material to porosity with the Bruggeman equation, $\tau=1/\sqrt{\varepsilon}$, like we will also do in the electrode. Furthermore, $D_{i}$ is the ion diffusion coefficient in free solution, $z_{i}$ the valency of the ion, and  $\phi$ is the dimensionless electric potential. %in the spacers. 
Dimensions of this potential can be restored by multiplication with the thermal voltage $V_\text{T}=RT/F=k\s{B}T/e$. 

Mass conservation describes the change of ion concentration with time at any position in these channels according to
\begin{equation}\label{eq:substantial_conservation}
\varepsilon_\text{s}\dfrac{\partial c_i}{\partial t}=-\nabla\cdot\bm{J_{i}}
\end{equation}
which can be combined with Eq.~\eqref{eq:nernst_planck_vector} and further simplified by assuming that
\vspace{-2.5mm}
\begin{itemize}
    \item Water flow in the spacer channels only has a component in the $y$--direction parallel to the membrane and electrode (see Fig. 1); hence $v_\text{x}=0$, and
    \item Diffusion and electro-migration are only considered in the opposite, $x$--, direction (which is perpendicular to the $y$--direction of water flow).
\end{itemize}
\vspace{-2.5mm}
With these assumptions, the balance for each ionic species in the spacer channel becomes~\cite{Sonin_Probstein_1968}
\begin{equation}\label{eq:conservation_spacer}
\varepsilon_\text{s} \frac{\partial c_i}{\partial t}= -v_\text{y}\frac{\partial c_i}{\partial y} + \frac{\varepsilon_\text{s}}{\tau_\text{s}}D_{i} \left\{ \frac{\partial^2 c_{i}}{\partial x^2} + z_{i}\frac{\partial}{\partial x}\left(c_{i}\frac{\partial \phi}{\partial x}\right) \right\} 
  \end{equation}
which is combined with local charge neutrality, $\sum_{i} z_{i} c_{i} =0$ and solved for each ionic species. For fluid velocity, we assume plug flow, thus $v_\text{y}$ is independent of $x$. 

The description of ion transport in the spacer channels acts as the basis for understanding the transport of ions in the porous electrodes, which results %for each ion 
in the ion mass balance 
\begin{equation}\label{eq:conservation_electrode}
   \varepsilon_\text{e} \frac{\partial c_{i}}{\partial t}=  \frac{\varepsilon_\text{e}}{\tau_\text{e}}D_{i} \left\{\frac{\partial^2 c_{i}}{\partial x^2} + z_{i}\frac{\partial}{\partial x}\left(c_{i}\frac{\partial \phi}{\partial x}\right) \right\} - a{J}_{\text{int},i} 
  \end{equation}
where $\varepsilon_\text{e}$ is electrode porosity (volume fraction of electrode filled with electrolyte), and $\tau_\text{e}$ is electrode tortuosity. In the electrode the fluid velocity is set to zero. Similar to the spacer channel, electroneutrality is assumed in the electrolyte pore phase in the electrodes. In the model presented here, this corresponds to [Na$^+$] + [Cl$^-$]. The new addition in Eq.~\eqref{eq:conservation_electrode}, $a{J}_{\text{int},i}$, is the ion flux from electrolyte into the intercalation material (redox-active nanoparticles), and is defined per unit area $a$, which is the area of the particles per unit volume of total electrode. %Parameter $s$ is the selectivity of the intercalation host particles and 
For a perfectly cation-selective material, we have for all anions ${J}_{\text{int},i}=0$, while for the cations, ${J}_{\text{int},i}$ follows from a balance over the nanoparticles. %the value $s=+1$ for a cation and $s=0$ for an anion. 
Assuming only a single cation to intercalate in the intercalation material, a cation mass balance for the nanoparticles is given by
\begin{equation}\label{eq:ihc}
c_\text{max}\varepsilon_\text{im}\frac{\partial \vartheta}{\partial t}=a{J_\text{int}}
\end{equation}
where we leave out subscript $i$, and where $c_\text{max}$ is the maximum possible concentration of cations in the intercalation material, $\varepsilon_\text{im}$ is the volume fraction of the intercalation material (as a fraction of the total electrode volume), and $\vartheta$ is the average cation intercalation degree (average fraction of active sites in the host particles occupied by a cation).

Finally, local chemical equilibrium is assumed between cations in the intercalation material and in the electrolyte. This chemical equilibrium is described using the Frumkin isotherm for intercalation~\cite{Conway_Gileadi,West_1982,Porada_Smith}
\begin{equation} \label{eq:frumkin}
\phi_\text{ecm}-\phi_\infty=\mu_\text{}^\dagger-\ln\frac{\vartheta_\text{}}{1-\vartheta_\text{}} + \ln\frac{c_{+}}{c_\text{ref}} - g \left( \vartheta_\text{} - \nicefrac{1}{2}  \right)
\end{equation}
where $\phi_\text{ecm}$ is the potential of the electron-conducting material (carbon in most cases), and $\phi$ is the potential in the nearby electrolyte phase, which is the same $\phi$ as used in Eq.~\eqref{eq:conservation_electrode}. Furthermore, $c_{+}$ is the concentration of cations there, $c_\text{ref}$ is a reference concentration, and $\mu_\text{}^\dagger$ and $g_\text{}$ are constant factors dependent on the type of cation, where $g_\text{}$ can be considered as an inter-cation repulsion energy. The Frumkin isotherm fits very well equilibrium data for electrode potential versus charge for electrodes containing the intercalation material NiHCF, a Prussian Blue analogue~\cite{Porada_Smith}.  As noted above, the Frumkin isotherm is equivalent to the regular solution model for the Gibbs free energy of mixing of intercalated ions and vacancies, which has also been successfully used in phase-field models of Li-ion batteries~\cite{Bazant_2013,Lai_2010,Smith_Bazant_2017}.

In the electron-conducting phase in the electrode, the electronic resistance is considered to be zero. This implies that the potential here, $\phi_\text{ecm}$, does not vary with position. However, it is different for cathode and anode, and it varies in time. Thus, the voltage as measured between anode and cathode, the cell voltage, is given by $V_\text{cell}=V_\text{T}\cdot \left(\phi_\text{ecm,A}-\phi_\text{ecm,C}  \right)$ where ``A'' and ``C'' refer to anode and cathode.

Finally, ion transport in the AEM separating the two halves of the cell must be described. As in the other regions, transport of ions in the AEM is described by the NP equation~\eqref{eq:nernst_planck_vector}. Ion transport by advection is neglected in the membrane. Similar to the electrodes and the spacer channels, it is  assumed that diffusion and migration occur only in the  $x-$direction across the membrane of thickness $\delta_\text{m}$. Accumulation of mass is neglected in the membrane and as a consequence, in the membrane the flux of each ion remains invariant with position $x$. An equal diffusion coefficient $D_\text{m}$, in which the term $\varepsilon/\tau$ is included, is assumed for all ions in the membrane, and we define a transfer coefficient $k_\text{m}=D_\text{m}/{\delta_\text{m}}$. The membrane is furthermore defined by the fixed charge density, $\omega X$, which is  a positive number for an AEM. A typical value for a commercial membrane is of the order of $X=4$ M. This is a fixed charge concentration per unit aqueous phase in the membrane. For a binary monovalent electrolyte, charge neutrality in the membrane is given by $c_\text{+,m} - c_\text{-,m} + \omega X =0$ and a total ion concentration in the membrane is defined as  $c_\text{T,m}=c_\text{+,m} + c_\text{-,m}$.  The total ion flux is $J_\text{ions,m}=J_\text{+,m}+J_\text{-,m}$ and the charge flux (ion current density) is $J_\text{ch,m}=J_\text{+,m}-J_\text{-,m}$. 

When the fixed charge density in the membrane is much higher than the salinity outside the membrane, the ion concentrations and the electrical potential, $\phi$, can be assumed to be varying linearly with position inside the membrane~\cite{Mubita_EA}. As a consequence, the fluxes $J_\text{ions,m}$ and $J_\text{ch,m}$  become
\begin{align}
\begin{split}
J_\text{ions,m}&=-{k_\text{m}} \left( \Delta c_\text{T,m} - \omega X \Delta \phi_\text{m} \right)    \\
J_\text{ch,m}&=-{k_\text{m}}\left<c_\text{T,m} \right>\Delta \phi_\text{m}
\end{split}
\end{align}
where $\left< ... \right>$ is an average between values at the extreme ends of the membrane, and $\Delta$ refers to a difference between these positions (with fluxes defined from left to right, $\Delta$ is defined as ``right'' minus ``left''). The total ion concentration on each side of the membrane is related to the salt concentration just outside (i.e., in the spacer channel, at a position right next to the membrane), $c^*$, as
\begin{equation}
c_\text{T,m}=\sqrt{X^2+\left(2 c^*\right)^2}
\end{equation}
while the Donnan potential (change in electrical potential) across each membrane edge is given by
\begin{equation}
\Delta \phi_\text{D}=\sinh^{-1}\left({\omega X}/{2 c^*}\right).
\end{equation}

\section{Results and Discussion} \label{s:results}

The calculations for the CDI cell are set up for an electrolyte with a single cation, Na$^+$, that can intercalate into the redox-active nanoparticles under the influence of an applied electrical current. As a specific geometrical example of an electrochemical desalination cell, we consider CDI with porous electrodes containing intercalation particles where the electrode has the spacer channel on one side and the ``current collector'' on the other side. Here, electronic current is injected in the electrode. The spacer channels are positioned next to the AEM which is in the middle of the cell, as shown in Fig.~\ref{fig:schematic_cell}. The electrolyte, a NaCl solution, flows through the spacer channel and both ions  diffuse in and out of the electrodes, while Cl$^-$-ions (and some Na$^+$-ions) also diffuse through the membrane. Note that also other designs are possible: in ref.~\cite{Smith_Dmello} a cell design was theoretically analyzed with the water flowing through the electrodes, and in ref.~\cite{Porada_Smith} the spacer channel was located on the other side of the electrodes (on the outsides of the cell). 

\subsection{Numerical aspects}

The CDI cell model for our geometry is solved numerically, based on the conservation equations (\ref{eq:conservation_spacer}--\ref{eq:ihc}). These equations are used for both ions, for spacer and electrode, along with the Frumkin isotherm (\ref{eq:frumkin}) and the equations for the membrane. In $x$-direction (left-right direction in Fig. 1), discretization of the partial differential equations in each domain is done using a large number of nodes (>20 in each domain), but in $y$--direction we use only use a single node and an implicit Euler scheme (backward Euler). Because of this choice it is as if in the flow direction each flow line is one (very thin) ``stirred tank'', an approach successfully applied for CDI in ref.~\cite{Dykstra_WR} and for electrodialysis in ref.~\cite{Tedesco_JMS}. For the low desalination degrees as obtained in the present work, discretization in the $y$--direction into more nodes, will not significantly change the results of the calculations. Note that in case of ``one node in $y$--direction,'' the geometry of how inlets and outlets of the two cell compartments are oriented relative to one another (e.g., co-current vs counter-current) does not matter; this only starts to influence the calculation outcome with a higher discretization in $y$--direction. % that the orientation of flow direction must be defined, because it influences the calculation. As a side-note, it is interesting to note that (in our model, which neglects diffusion, migration and dispersion in $y$--direction) whether the flow compartment is long and thin or more square-like, does not play any role in case we have perfect co-current flow or perfect counter-current flow with parallel flowlines. (For circular electrodes, with flow in radial direction, in some cases we also have the same calculation outcome.) 
Another numerical aspect of our model is as follows: with more than one node in $y$--direction, constant current operation, where a certain current $I$ is made to flow from anode to cathode, requires solving all equations on each node in $y$--direction simultaneously, because current $I$ will not  %necessarily distribute evenly over each part of the cell (i.e., it is not 
evenly distributed over all $y$--coordinates. However, when a boundary condition of constant cell voltage is used, then it is actually possible to solve each $y$--node ``sequentially'', i.e., one after the other. Especially for steady-state operation in electrodialysis, this is particularly useful. %(As a detail, this is only possible when an external electric resistance that is ``shared'' among all $y$--nodes can be neglected.)
Finally, note that because we have neglected diffusion and migration in $y$--direction, we do not have to deal with various numerical complications at the inlet and outlet of the cell, such as ionic currents straying outside the cell, or the %difficult 
choice of correct boundary conditions in advection-diffusion problems, where the concentration decay in $y$--direction can propagate upward into feed tubes, see e.g. ref.~\cite{Guyes_2017}.

For our calculations, parameters are based on experiments in ref.~\cite{Porada_Smith}, resulting in: $c_\text{salt,inflow}=10$ mM, $\varepsilon\s{s}=1.00$, $\varepsilon\s{e}=0.30$, $\varepsilon\s{im}=0.50$, $c\s{max}=4.4$ M, cell area $A=36$ cm$^2$, electrode thickness $\delta\s{e}=250$ $\mu$m, spacer thickness $\delta\s{s}=700$ $\mu$m, and water flow rate per channel $\Phi_\text{v}=5.0$ mL/min, which results in a residence time of (water in) each spacer channel of $A\delta\s{s}/\Phi\s{v}=30$ s. Diffusion coefficients of Na$^+$ and Cl$^-$ are $1.33\times 10^{-9}$ m$^{2}$/s and $2.00\times 10^{-9}$ m$^{2}$/s, respectively. For the membrane, the transfer coefficient is set to the value $k\s{m}=0.55$ $\mu$m/s (based on taking the average of $D_{\text{Na}^+}$ and $D_{\text{Cl}^-}$, reducing that by a factor of 20, and implementing a membrane thickness of $\delta\s{m}=150$ $\mu$m). Membrane charge density is $X=4.0$ M ($\omega=+1$ for an AEM). Operation is always with constant current at a value of $I=\pm 10$ mA. With this current running for one hour, this implies that the average intercalation degree in a certain electrode then changes by about $\Delta\vartheta=0.20$, because $A\delta\s{e}F \varepsilon\s{im}c\s{max}\Delta\vartheta=I \Delta t$ where $\Delta t$ is the time period of charging. The Temkin parameter of relevance is the inter-cation repulsion $g$, for which we use $g=3.50$ (dimensionally, 90 mV). The factor $\mu^\dagger$ does not influence any calculation output and neither does the choice of $c_\text{ref}$.

\subsection{Initial charging and discharge}

In all cases we start the calculation with both electrodes charged half-way, i.e., $\vartheta=0.5$, without gradients in $\vartheta$ across the electrode. Then we charge one electrode relative to the other for a period of 1 hr, resulting in one electrode reaching an average $\vartheta\s{avg}=0.30$ and the other $\vartheta\s{avg}=0.70$. Then we discharge (i.e., reverse the direction of current) for the same duration of 1 hr, after which each electrode is back to $\vartheta\s{avg}=0.50$. These are average values of $\vartheta$, with the profiles in $\vartheta$ across each electrode as presented in Fig.~2c for different moments during the first hour of charging, and in Fig.~2d for various times during discharge. The salt concentration profiles during these periods are presented in Figs. 2a and 2b. This initial cycle of charge and discharge is also shown as the first 2 hr in Fig. 3, as well as the first 2 hr in Fig. 4.  

\subsection{Two modes of operation of charge/discharge cycles}

After that time (from $t=2$ hr onward), two distinct operational modes are considered. In Mode~1, at $t=2$ hr we again reverse the current direction and go back to the situation we had at $t=1$~hr, at least in terms of $\vartheta\s{avg}$. This cycle is then repeated with the current reversed every hour. Mode~1 is depicted in Fig. 3a,c and in Fig 4. Instead, in Mode~2, at $t=2$~hr, nothing is changed, and the current continues such that after another hour, at $t=3$~hr, we reach a situation that the values of $\vartheta\s{avg}$ are now reversed compared to the situation at $t=1$ hr, i.e., the electrode that had $\vartheta\s{avg}=0.70$ at $t= 1$ hr, now, at $t=3$ hr, has $\vartheta\s{avg}=0.30$ (and vice-versa). Only now, at time $t=3$~hr, is the current reversed, and a period of charging starts that in 2 hr will bring us back to the other end of the cycle (so that in terms of $\vartheta_\text{avg}$, the cell is the same at $t=5$ hr as at $t=1$~hr). The main advantage of Mode~2 operation is that (when the limit cycle is reached, and for a perfectly symmetric system) operation goes from a cell voltage of $+V$ at one end of the cycle, to a voltage of $-V$ at the other end, see Fig. 4. This symmetry is not found for operation according to Mode~1. Thus Mode 2-operation is very easily established experimentally by setting a value of $\pm V$ at which the cell reverses the direction of current.

Interestingly, in this Mode 2 of operation where current is reversed when the cell voltage is $\pm V$, each ``half-cycle'' is now exactly the same as that in the next half-cycle, with the only difference being that what happens in electrode 1 during a certain half-cycle, now happens in electrode 2 in the next. This can be observed in Fig. 3d, where the red dashed line during period 1--3 hr, is almost the same as the blue solid line in the period 3--5 hr, and the same for the blue line in the period 1--3 hr which can be compared with the red dashed line in the period 3--5 hr. In Fig. 3, the behavior is not yet exactly the same because the limit cycle or ``dynamic steady-state'' was not yet reached after 1 hr. Note that for operation in this Mode 2, it is impossible to find an objective criterion about which half of the cycle is to be called ``charge'' and which ``discharge''. There is no criterion because both halves of the cycle are exactly identical in terms of desalination performance, energy consumption, etc. Note that this exact symmetry only occurs when for both halves of the cell all operational setting are the same such as water flow rate and electrode mass. Interestingly, it is not necessary for both electrodes to have the same initial $\vartheta$ at time zero when we operate by fixing the end-voltages of each step to $\pm V$.

\begin{figure}[ht]
\centering
\includegraphics[width=0.5\textwidth]{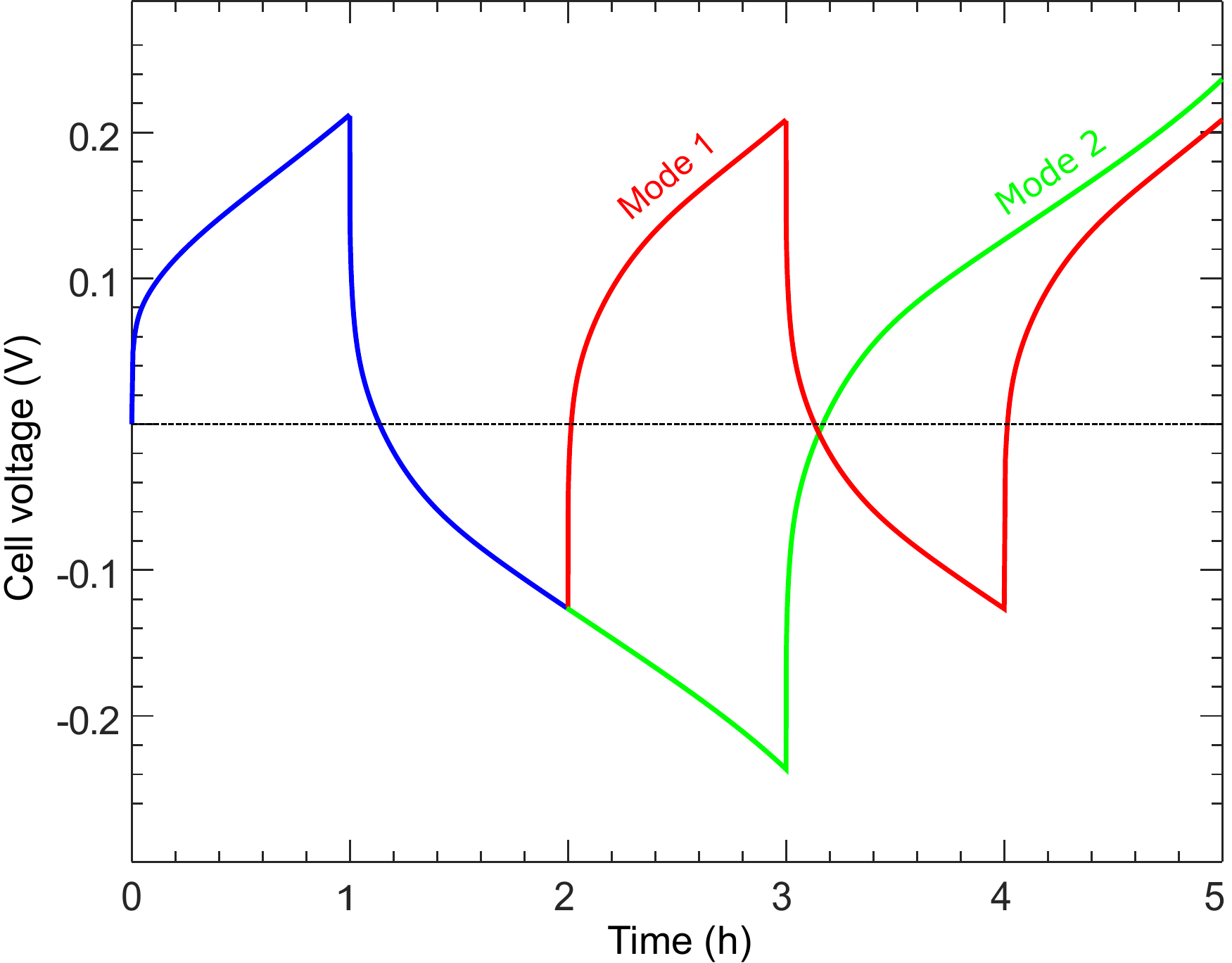}
\caption{Cell voltage for Mode 1 and Mode 2 of operation. The cell voltage changes suddenly when current changes direction. The blue curve shows the first cycle, while red and green refer to operation in Mode 1 and Mode 2, respectively.}
\label{fig:voltage_vs_time}
\end{figure}

Finally we show in Fig.~\ref{fig:voltage_vs_time} calculation results for the development in time of cell voltage  for the two modes of operation. These voltage ($V$)--time ($t$) plots are important as they fully describe the electrical energy input in the process: for a constant current of $\pm I$, the electrical energy input is proportional to the area under the $V$--$t$ curve (and delineated by the line $V\s{cell}=0$). As Fig. 4 shows, after reversal of the current direction, there is also a small ``area'' which corresponds to energy ``output'' (e.g. from 1 to 1.15 hr), i.e., where energy recovery is in theory possible~\cite{Hawks_2018}, after which a much longer period follows where energy must be invested to run the cell (from 1.15 to 3 hr for Mode 2).
         
\section{Conclusions} 
In conclusion, in this work we combined porous electrode theory for electrodes with redox-active intercalation materials with a description of spacer channels and a membrane, to describe an electrochemical system for capacitive deionization (CDI) with two intercalation electrodes that are chemically the same. The CDI cell design has an intrinsic symmetry as it is composed of two identical porous electrodes, both interfacing a spacer channel,  separated by one anion-exchange membrane. We considered a single-salt solution, and calculated transport of ions by diffusion, migration and advection in two dimensions. Transport of ions in the pores of the electrodes was modeled based on the Nernst-Planck equation. Thermodynamics of cation adsorption from electrolyte into the intercalation material was described by the Frumkin isotherm (regular solution theory), which factors in the lattice occupancy, inter-cation repulsion, electrolyte salinity and electrode potential. Homogeneous adsorption of ions was assumed, i.e., without phase separation within the nanoparticles. Including of this effect may be of importance for certain material chemistries. The required model extensions can be based on approaches discussed in refs.~\cite{Ferguson_2012,Bazant_2013,Smith_Bazant_2017}. 

%Both parts of the model, transport and thermodynamics, can also be extended to consider a mixed ionic system with multiple cations and anions. 
In our model, a limitation in transport of ions within the intercalation material was neglected, just as was a resistance for flow of electronic current across the electrode. The model describes the development in time of salt concentration, intercalation degree and cell voltage, parameters important for the description of performance of a CDI desalination cell. This work intends to provide a platform upon which future models can be constructed for more realistic water sources with ionic mixtures and for other geometries of desalination systems involving porous electrodes with redox-active intercalation materials. 

\section*{Acknowledgments}

This work was supported by the European Union's Horizon 2020 research and innovation programme %Research Council 
(ERC Consolidator Grant, agreement no. 682444) %, E-motion, PI:LCPMDS) 
and was performed in the cooperation framework of Wetsus, European Centre of Excellence for Sustainable Water Technology (www.wetsus.eu). Wetsus is co-funded by the Dutch Ministry of Economic Affairs and Ministry of Infrastructure and Environment, the Province of Frysl{\^a}n, and the Northern Netherlands Provinces. %We thank members of the research theme Capacitive Deionization at Wetsus for useful discussions.  

\end{document}